\documentclass[aps,prb,twocolumn,fleqn,eqsecnum,floatfix]{revtex4}

\usepackage{amssymb}
\usepackage{amsmath}
\usepackage{graphicx}

\newcommand{\Tr}{\operatorname{Tr}}

\begin{document}

\title{Non-ergodic dynamics of the extended anisotropic Heisenberg chain}

\author{Evgeny Plekhanov}
\email[E-mail: ]{plekhanoff@sa.infn.it}
\author{Adolfo Avella}
\email[E-mail: ]{avella@sa.infn.it}
\author{Ferdinando Mancini}
\email[E-mail: ]{mancini@sa.infn.it} \homepage[Group Homepage:
]{http://scs.sa.infn.it}

\affiliation{Dipartimento di Fisica ``E.R. Caianiello'' - Unit\`{a}
CNISM di Salerno \\
Universit\`{a} degli Studi di Salerno, I-84081 Baronissi (SA),
Italy}

\begin{abstract}
The issue of ergodicity is often underestimated. The presence of
zero-frequency excitations in bosonic Green's functions determine
the appearance of zero-frequency momentum-dependent quantities in
correlation functions. The implicit dependence of matrix elements
make such quantities also relevant in the computation of
susceptibilities. Consequently, the correct determination of these
quantities is of great relevance and the well-established practice
of fixing them by assuming the ergodicity of the dynamics is quite
questionable as it is not justifiable a priori by no means. In this
manuscript, we have investigated the ergodicity of the dynamics of
the $z$-component of the spin in the 1D Heisenberg model with
anisotropic nearest-neighbor and isotropic next-nearest-neighbor
interactions. We have obtained the zero-temperature phase diagram in
the thermodynamic limit by extrapolating Exact and Lanczos
diagonalization results computed on chains with sizes $L = 6 \div
26$. Two distinct non-ergodic regions have been found: one for
$J^\prime/J_z \lesssim 0.3$ and $|J_\perp|/J_z < 1$ and another for
$J^\prime/J_z \lesssim 0.25$ and $|J_\perp|/J_z = 1$. On the
contrary, finite-size scaling of $T \neq 0$ results, obtained by
means of Exact diagonalization on chains with sizes $L = 4 \div 18$,
indicates an ergodic behavior of dynamics in the whole range of
parameters.
\end{abstract}

\maketitle

\section{Introduction}\label{intro}

In order to study a physical system (i.e., to analyze its properties
and its responses to external perturbations), we have to compute
correlation functions $C$ and susceptibilities $\chi$ (retarded
Green's functions) of any operator $\psi$ relevant to the dynamics
under investigation
\begin{align}
&C(i,j)=\left\langle\psi(i)\psi^{\dagger}(j)\right\rangle\\
&\chi(i,j)= -\mathrm{i}
\theta(t_i-t_j)\left\langle\left[\psi(i),\psi^{\dagger}(j)\right]\right\rangle,
\end{align}
where $\langle \cdots \rangle$ stands for the average in some
statistical ensemble, $i=({\bf i},t)$ for both spatial ${\bf i}$ and
temporal $t$ coordinates. The operator $\psi$ is taken in the
Heisenberg picture.

In principle, one can compute the complete set of eigenstates
$|n\rangle$ and eigenvalues $E_{n}$ of the Hamiltonian describing
the system under study. This knowledge makes possible to compute the
Fourier transform in frequency of any correlation function as
follows
\begin{multline}
C({\bf i},{\bf j},\omega) = 2\pi\delta(\omega) \Gamma ({\bf i},{\bf j}) + \nonumber \\
{\frac{2\pi}{Z}} \sum_{ \stackrel{n,m}{E_{n} \neq E_{m}}} {\rm
e}^{-\beta E_{n}} \langle n | \psi({\bf i}) | m \rangle \langle m |
\psi^{\dagger}({\bf j}) | n \rangle \delta \left[ \omega + \left(
E_{n} - E_{m} \right) \right]
\end{multline}
where the {\em zero-frequency function} $\Gamma({\bf i},{\bf j})$ is
defined as \cite{Mancini_00,Mancini_04}
\begin{equation}\label{Gammadef}
\Gamma({\bf i},{\bf
j})=\frac{1}{Z}\sum_{\stackrel{n,m}{E_{n}=E_{m}}} {\rm e}^{-\beta
E_{n}} \langle n | \psi({\bf i}) | m \rangle \langle m |
\psi^{\dagger}({\bf j}) | n \rangle .
\end{equation}
$\Gamma$ explicitly appears in the expression of the correlation
function when the field $\psi$ is boson-like, that is, $\psi$ is
constituted of an even number of original electronic (i.e.,
fermionic) operators \cite{Mancini_00,Mancini_04}. Hereafter, we
focus on such a case.

According to the well-known relations existing between casual ($C$),
retarded ($R$) Green's functions and correlation functions
\begin{align}
&\Re [G^{R} ({\bf k}, \omega)]=\Re [G^{C} ({\bf k},
\omega)] \\
&\Im [G^{R} ({\bf k}, \omega)]= \tanh \left ( {{{\beta
\omega } \over 2}} \right)\Im [G^{C} ({\bf k}, \omega)] \\
&C({\bf k}, \omega)=-\left[ {1+\tanh \left ( {{{\beta \omega } \over
2}} \right)} \right]\Im [G^{C} ({\bf k}, \omega)]
\end{align}
the zero-frequency excitations do not contribute explicitly to the
imaginary part of the retarded Green's functions and, consequently,
$\Gamma$ does not explicitly appear in the expressions of
susceptibilities. At any rate, susceptibilities retain an implicit
dependence on $\Gamma$ through the matrix elements $\langle n |
\psi({\bf i}) | m \rangle$. Then, the right procedure to compute
both correlation functions and susceptibilities is clearly the one
that starts from the causal Green's function, which is the only
Green's function that explicitly depends on $\Gamma$.

It is worth noticing that the value of $\Gamma$ dramatically affects
the values of directly measurable quantities (e.g., compressibility,
specific heat, magnetic susceptibility, \ldots) through the values
of correlation functions and susceptibilities. According to this,
whenever it is possible, $\Gamma$ should be exactly calculated case
by case.

If we do not have access to the complete set of eigenstates and
eigenvalues of the system, which is the rule in the most interesting
cases, we have to compute correlation functions and susceptibilities
within some, often approximated, analytical framework. Now, since no
analytical tool can easily determine $\Gamma$ (e.g., the equations
of motion cannot be used to fix $\Gamma$ as it is constant in time),
one usually assumes the ergodicity of the dynamics of $\psi$ and
simply substitutes $\Gamma$ by its ergodic value:
\begin{equation}
\Gamma^{erg}({\bf i},{\bf j})=\langle \psi({\bf i}) \rangle \langle
\psi^{\dagger}({\bf j}) \rangle.
\end{equation}
Unfortunately, this procedure cannot be justified a priori (i.e.,
without computing $\Gamma$ through its definition (\ref{Gammadef}))
by absolutely no means. The existence of just one integral of motion
divides the phase space into separate subspaces not connected by the
dynamics. This latter, in turn, becomes non ergodic: time averages
give different results with respect to ensemble averages.

The lack of ergodicity has sizable effects. One of its
manifestations is the well-known difference between the static
isolated (or Kubo) susceptibility and the isothermal one
\cite{Suzuki_71}. The dynamics of an operator $\psi$, in a system
described by the Hamiltonian $H$, is said to be ergodic if
\cite{Suzuki_71}
\begin{equation} \label{erg-def}
\lim_{\tau \to \infty} \frac1\tau \int_0^\tau \langle
\psi\psi^{\dagger}(t) \rangle \, dt = \langle \psi \rangle \langle
\psi^{\dagger} \rangle,
\end{equation}
where $\langle \cdots \rangle$ stands, for instance, for the
canonical average at temperature $T$
\begin{equation} \label{therm-av}
\langle \psi \rangle = \frac1Z \Tr(\psi e^{-\beta H}),
\end{equation}
with $\beta=\frac1T$ and $Z=\Tr(e^{-\beta H})$. That is, the
dynamics of $\psi$ is ergodic if, during its time evolution, it has
non-zero matrix elements only between states within a zero-volume
region of the phase space of the system. Only in this case, the
equilibrium averages (l.h.s. of (\ref{erg-def})) are equal to the
ensemble averages (r.h.s. of (\ref{erg-def})), which are much easier
to compute. Now, one can show \cite{Morita_69} that the difference
between the static (or Kubo) susceptibility
\begin{equation}\label{Kubo}
\begin{split}
\chi(0)&=\lim_{\omega \to 0} \mathcal{F}[ -\mathrm{i} \, \theta(t)
\langle [ \psi, \psi^{\dagger}(t)] \rangle ]\\
&=\int_0^\beta \langle \psi \psi^{\dagger}(\mathrm{i}\lambda)
\rangle \, d\lambda - \beta \lim_{\tau \to \infty} \frac1\tau
\int_0^\tau \langle \psi \psi^{\dagger}(t) \rangle \, dt,
\end{split}
\end{equation}
and the isothermal susceptibility
\begin{equation}
\chi^{T} =  \left. \frac{ \delta \langle \psi \rangle }{ \delta
\mathcal{H}} \right|_{\mathcal{H} \to 0}= \int_0^\beta \langle \psi
\psi^{\dagger}(\mathrm{i}\lambda) \rangle \, d\lambda - \beta
\langle \psi \rangle \langle \psi^{\dagger} \rangle ,
\end{equation}
is just proportional to the difference between the l.h.s. and the
r.h.s. of (\ref{erg-def})
\begin{equation} \label{non-erg1}
\chi^{T} - \chi(0) = \beta \left( \lim_{\tau \to \infty} \frac1\tau
\int_0^\tau \langle \psi \psi^{\dagger}(t) \rangle \, dt - \langle
\psi \rangle \langle \psi^{\dagger} \rangle \right).
\end{equation}
In the above, $\mathcal{F}$ stands for the Fourier transform and
$\mathcal{H}$ for the external field coupled to $\psi$ (e.g., an
external magnetic field in the case of magnetization). The second
line of (\ref{Kubo}) comes from the relation between susceptibility
and relaxation function. $\Gamma$ (as defined in (\ref{Gammadef}))
is exactly equal to the l.h.s. of (\ref{erg-def}).

Thus, it is clear that issue of ergodicity is relevant to linear
response too. It is also worth noting that non-ergodicity at zero
temperature is just the result of a degeneracy in the ground state,
which would be lifted by a vanishing $\psi$-$\mathcal{H}$ coupling
in the Hamiltonian. For instance, the reader can imagine a
ferromagnetic ground state with finite spontaneous magnetization
driven by a vanishing external magnetic field.

Many systems exhibit non-ergodic dynamics. Over the last years, we
have been studying \cite{Mancini_00,Mancini_05b} some of them
analytically by means of the Composite Operator Method
\cite{Mancini_00,Mancini_04}: two-site Hubbard model, three-site
Heisenberg model, tight-binding model in magnetic field,
double-exchange model, extended Hubbard model in the ionic limit.
Recently, we have started an investigation of such models also by
means of numerical techniques: Lanczos and Exact Diagonalization.
Along this latter line, in the present paper, we calculate $\Gamma$
for the $z$-component of the spin at site $\mathbf{i}$,
$S^z_\mathbf{i}$, in the one-dimensional anisotropic extended
Heisenberg model (see next section) and show that it takes
non-ergodic values in some regions of the parameter space of the
model both for finite sizes and in the bulk limit. This analysis is
the natural continuation of that reported in
Ref.~\onlinecite{Bak_02a}, where we have analyzed the standard
(isotropic and non-extended) Heisenberg model at zero temperature
and found a non-ergodic behavior of $\Gamma$.

\section{Definitions and Method} \label{model}

We have studied the ergodicity of dynamics of the operator
$S^z_{\mathbf{i}}$ in the 1D anisotropic extended Heisenberg model
described by the following Hamiltonian:
\begin{multline} \label{ham}
H = -J_z \sum_\mathbf{i} S^z_\mathbf{i} S^z_{\mathbf{i}+1} \\
+J_\perp \sum_\mathbf{i} ( S^x_\mathbf{i} S^x_{\mathbf{i}+1} +
S^y_\mathbf{i} S^y_{\mathbf{i}+1}) +J^\prime \sum_\mathbf{i}
\mathbf{S}_\mathbf{i} \mathbf{S}_{\mathbf{i}+2},
\end{multline}
where $S^x_\mathbf{i}$, $S^y_\mathbf{i}$ and $S^z_\mathbf{i}$ are
the $x$, $y$ and $z$ components of the spin-$1/2$ at site
$\mathbf{i}$, respectively. The model~(\ref{ham}) is taken on a
linear chain with periodic boundary conditions. We take the
interaction term parameterized with $J_z$ ferromagnetic ($J_z>0$)
and the next-nearest-neighbor interaction term, which is
parameterized with $J^\prime$, isotropic. In order to frustrate
ferromagnetism, we have considered only the case with $J^\prime>0$,
that is, with an antiferromagnetic coupling between next-nearest
neighbors. According to this, only chains with even number of sites
have been studied in order to avoid topological frustration that
would be absent in the thermodynamic limit.  Since it is possible to
exactly map all results obtained for $J_\perp
> 0$ to those for $J_\perp < 0$ by means of a simple canonical
transformation, we have limited our study only to positive values of
$J_\perp$.

We have found some numerical
\cite{Tonegawa_87,Harada_88,Nomura_93,Nomura_94,Hirata_00} and few
analytical studies \cite{Haldane_82,Shastry_81} of the model
(\ref{ham}) in the literature, mainly in the antiferromagnetic state
($J_z<0$). Only one fixed point ($J_\perp=J_z$ and $J^\prime=J_z/2$)
is known to possess an analytic ground state \cite{Majumdar_70}. In
the rest of the phase diagram, the model (\ref{ham}) seems not to be
integrable \cite{Frahm_92,Cyr_96}.

We have numerically diagonalized the Hamiltonian (\ref{ham}) for
chains of size $L$ ranging between $6$ and $18$ by means of Exact
Diagonalization (ED) (divide and conquer algorithm) and for chains
of size $L$ ranging between $20$ and $26$ by means of Lanczos
Diagonalization (LD). We have systematically taken into account
translational symmetry and classified the eigenstates by the average
value of $S^z=\sum_\mathbf{i} S^z_{\mathbf{i}}$, which is a
conserved quantity. Whenever we have used ED, all eigenvalues and
eigenvectors of (\ref{ham}) have been calculated up to machine
precision and, therefore, we have been able to determine the exact
dynamics of the system for all temperatures. On the contrary, when
we have used LD, we have been limited to the zero-temperature case
since only the ground state can be considered exact in LD.

As already discussed above, the dynamics of an operator (e.g.,
$S^z_\mathbf{i}$) is ergodic whenever (\ref{erg-def}) is satisfied,
or equivalently, (\ref{Gammadef}) is equal to its \emph{ergodic}
value:
\begin{equation} \label{gamma-erg}
   \Gamma^{erg}=\langle S^z_\mathbf{i} \rangle^2
   =\frac{1}{Z^2}\sum_{n,m} e^{-\beta ( E_n + E_m ) }
   \langle n | S^z_\mathbf{i} | n \rangle
   \langle m | S^z_\mathbf{i} | m \rangle.
\end{equation}
The dynamics of a finite system is hardly ergodic, since
(\ref{Gammadef}) and (\ref{gamma-erg}) unlikely coincide. In the
thermodynamic limit, the sums in (\ref{Gammadef}) and
(\ref{gamma-erg}) become series and no conclusion can be drawn
\textsl{a priori}. Since we have diagonalized the Hamiltonian
(\ref{ham}) numerically (i.e., only for finite systems) and since $L
\to \infty$ is the most interesting case, we have analyzed our
results through finite-size scaling in order to speculate on the
properties of the bulk system.

\section{Zero-temperature Results} \label{t0}

If the ground state of (\ref{ham}) is $N$-fold degenerate then, at
$T=0$, (\ref{Gammadef}) and~(\ref{gamma-erg}) read as follows:
\begin{eqnarray*} \label{t0-av}
   \Gamma^{\phantom{erg}}&=&\frac{1}{N} \sum_{n, m=1}^{N}
   |\langle n | S^z_\mathbf{i} | m \rangle |^2 \\
   \Gamma^{erg}&=&\left ( \frac{1}{N} \sum_{n}^{N}
   \langle n | S^z_\mathbf{i} | n \rangle \right )^2,
\end{eqnarray*}
respectively.

Thanks to the translational invariance enjoined by the system
$\langle S^z_\mathbf{i} \rangle$ is independent of $\mathbf{i}$ and
proportional to the $z$-component of the total spin operator average
$\langle S^z_{tot} \rangle$. It is easy to show that, even if there
is a finite magnetic moment per site in any of such $N$ degenerate
ground states, $\langle S^z_\mathbf{i} \rangle$ at $T =0$ is always
zero in absence of magnetic field. Indeed, if a ground state with
non-zero $\langle S^z_\mathbf{i} \rangle = M$ exists, also another
ground state with $\langle S^z_\mathbf{i} \rangle = -M$ exists.
Thus, at zero temperature, $\Gamma^{erg}$ is always zero and the
only quantity of interest is $\Gamma$. A finite value of this latter
implies non-ergodicity. Obviously, if $N=1$ then both values
coincide. Therefore, a non-ergodic phase corresponds to degenerate
ground states with finite magnetization.

\begin{figure}
\includegraphics[width=8cm]{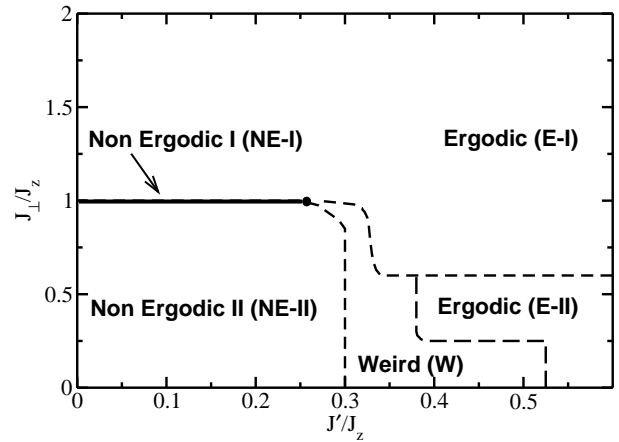}
\caption{Zero-temperature ergodicity phase diagram in the
$J^{\prime}-J_{\perp}$ plane. Due to the symmetry of the Hamiltonian
only the upper half is shown (see in the text). Only two ergodic
phases (E-I and E-II) have been found in the reported parameter
space. The others are either non-ergodic (NE-I and NE-II) or
impossible to conclusively analyze (W). The latter phase might
shrink to a transition line in the bulk limit.} \label{fig1}
\end{figure}

In the studied range of coupling constants (see Fig.~\ref{fig1}) we
have found two non-ergodic phases (NE-I and NE-II), two ergodic ones
(E-I and E-II) and a \emph{weird} phase (W). Our computational
facilities limit the range of chain sizes that we can analyze such
that we could not establish, by means of finite-size scaling,
whether the \emph{weird} phase (W) is ergodic or not. In the
non-ergodic phases (NE-I and NE-II), we were able not only to
perform the finite-size scaling, but also to write down an analytic
expression for $\Gamma$ as a function of the chain size $L$. The
\emph{weird} phase (W) has exhibited a strong dependence of the
ground state upon the particular values of the couplings. On the
contrary, the other phases exhibit ground states that are
independent of the particular values of the coupling constants.

\begin{figure*}[t]
\includegraphics[width=8cm]{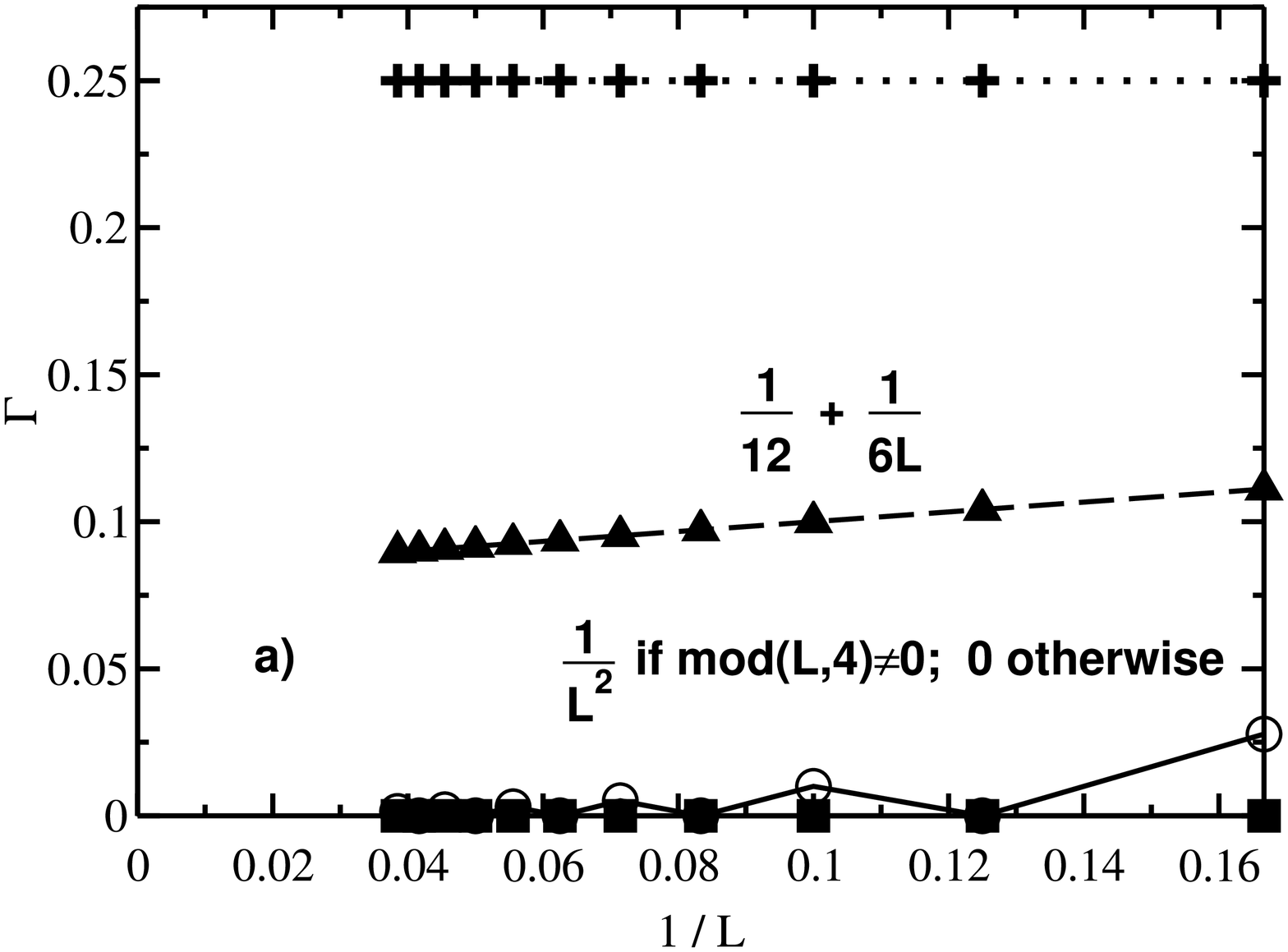}
\includegraphics[width=8cm]{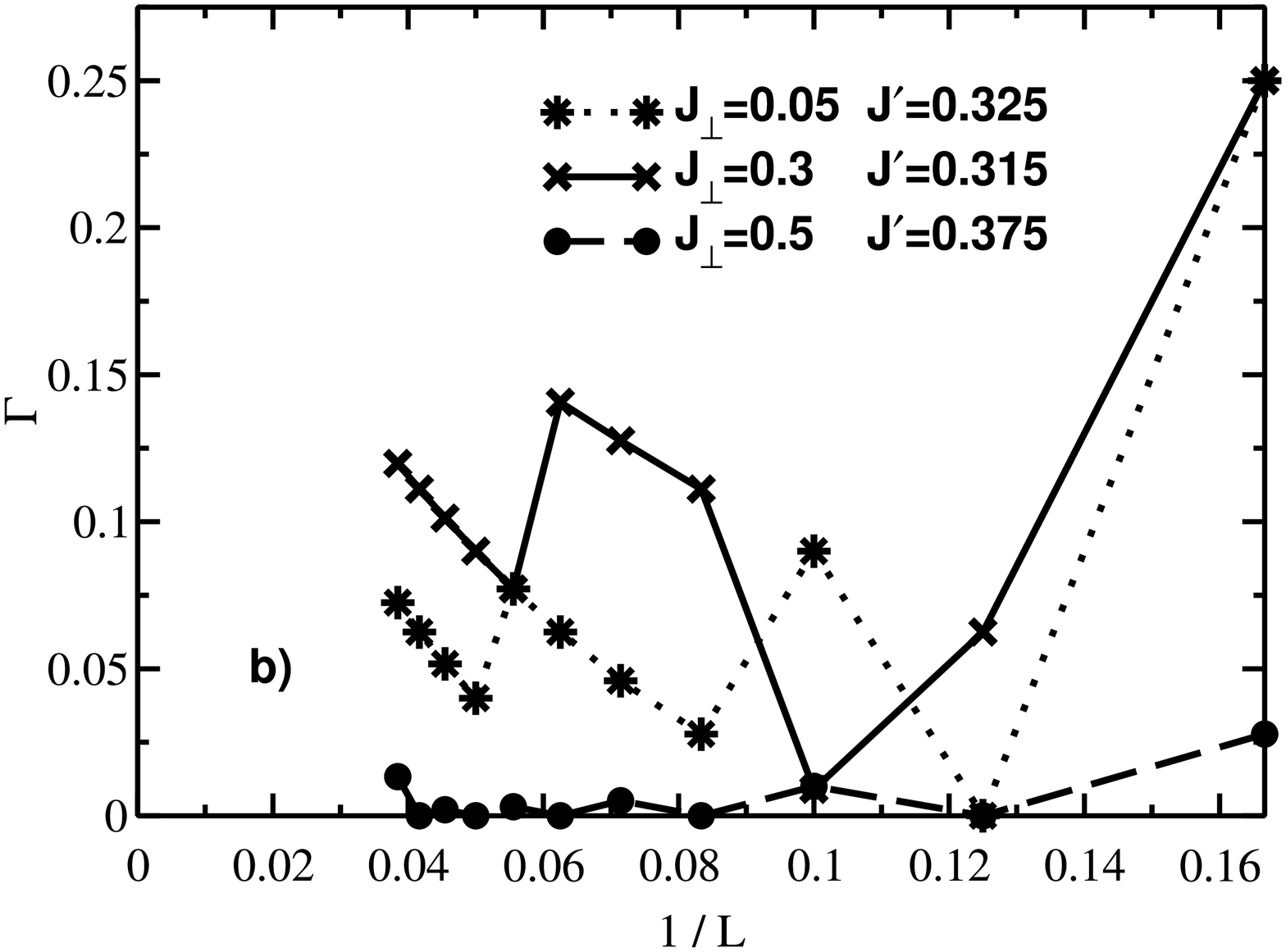}
\caption{\label{fig2} Finite-size scaling in the case of $T=0$ for
different points in the phase diagram of Fig.~\ref{fig1}. Symbols on
panel a): $+$ corresponds to (NE-II), $\blacktriangle$ corresponds
to (NE-I), $\blacksquare$ corresponds to (E-I) and $\bigcirc$ to
(E-II) regions of Fig.~\ref{fig1}, respectively. On panel b)
different examples from (W) region are shown. Hamiltonian couplings
are shown in the legend. All energies are expressed in units of
$J_z$.}
\end{figure*}

Ergodicity of the standard Heisenberg model ($J^\prime=0$ and
$J_\perp=J_z$) at $T=0$ has already been studied in
Ref.~\onlinecite{Bak_02a}: the dynamics is non-ergodic for
ferromagnetic coupling ($J_\perp=J_z<0$) as the system has a $L+1$
degenerate ground state
\begin{equation} \label{heis_ne}
\Gamma = \frac{1}{12} + \frac{1}{6L}.
\end{equation}
It is clear from (\ref{heis_ne}) that $\Gamma$ remains non-ergodic
also in the thermodynamic limit. This point ($J^\prime=0$ and
$J_\perp=J_z$) becomes a line in our phase diagram and is denoted as
NE-I (see Fig.~\ref{fig1}). In fact, the next-nearest-neighbor
interaction $J^\prime$ may frustrate ($J^\prime>0$) or favor
($J^\prime<0$) the ferromagnetism. In the latter case, the ground
state remains unchanged for any value of $J^\prime<0$. Therefore, we
expect the line denoting the phase NE-I to extend also to negative
$J^\prime$. If, on the contrary, $J^\prime$ is positive and large
enough to frustrate the system in such a way that the ground state
loses its ferromagnetic character, the ergodicity is restored. This
occurs at a finite critical $J^\prime \sim 0.25 J_z$. For values of
$J^\prime$ larger than the critical one, we find a non-degenerate
ground state with $\langle S^z_{tot} \rangle=0$.

If $J_\perp \neq J_z$ the rotational invariance is broken so that
states with the same $\langle {\bf S}_{tot}^2 \rangle$, but
different $\langle S_{tot}^z \rangle$, are not degenerate anymore.
In the non-ergodic region (NE-II) of the phase diagram (see
Fig.~\ref{fig1}), the ground state is just doubly degenerate (not
$L+1$ degenerate as in (NE-I)): one ground state corresponds to a
configuration with all spins \emph{up} and the other to a
configuration with all spins \emph{down}. Hence, the value of
$\Gamma$ in this phase is $1/4$ and does not depend neither on the
Hamiltonian couplings nor on the number of sites in the chain. It is
clear that also this phase extends to negative values of $J^\prime$.
This kind of ground state stands the frustration introduced by
next-nearest-neighbor interaction up to $J^\prime \sim 0.3J_z$ (see
Fig.~\ref{fig1}).

The ergodic region (E-I) of the phase diagram (see Fig.~\ref{fig1})
has $\Gamma=0$ for all sizes of the system and values of the
couplings: the unique ground state belongs to the sector with
$\langle S_{tot}^z \rangle = 0$. On the contrary, the other ergodic
phase (E-II) has non-zero values of $\Gamma$ for values of $L$ not
multiples of four. The ground state in this phase has average total
spin equal to one and, therefore, $\Gamma=1/L^2$. We obviously
conclude that (E-II) phase is ergodic in the thermodynamic limit.

The values of $\Gamma$ in these four phases (NE-I, NE-II, E-I and
E-II) exhibit perfect finite-size scaling as shown in
Fig.~\ref{fig2}a). This has allowed us to make definite statements
also in the thermodynamic limit.

The \emph{weird} phase (W) (see Fig.~\ref{fig1}) is characterized by
a quite strong size dependence, as shown in Fig.~\ref{fig2}b) where
a tentative finite-size scaling of $\Gamma$ in the different points
of the phase is presented. This region manifests a diverging
finite-size scaling within the range of sizes we were able to
handle. In this case, the behavior of $\Gamma$ as a function of $L$
strongly depends on the particular choice of the Hamiltonian
couplings and is highly non monotonous when increasing $L$,
according to the strong dependence on $L$ of $\langle S_{tot}^z
\rangle$ in the ground state. In this critical region the
eigenvalues of (\ref{ham}) present many level crossings, which means
that the maximum value of $L$ we were able to reach ($L_{max}=26$)
is not large enough to perform a sensible finite-size scaling
analysis. However, we expect that this phase becomes ergodic in the
thermodynamic limit, although still different from the ergodic
phases E-I and E-II.

We can summarize our findings in the thermodynamic limit at zero
temperature as follows:
\begin{equation}\label{gcase}
\Gamma=
\begin{cases}
\frac{1}{12} & \text{if $J_\perp= \pm J_z$ and $J^\prime \lesssim 0.25 J_z$} \\
\frac{1}{4} & \text{if $|J_\perp| < J_z$ and $J^\prime \lesssim 0.3 J_z$} \\
??? & \text{in the \emph{weird} phase (W) (see Fig.~\ref{fig1})}\\
0 & \text{otherwise}
\end{cases}
\end{equation}

\begin{figure*}[t]
\centerline{
\includegraphics[width=8cm]{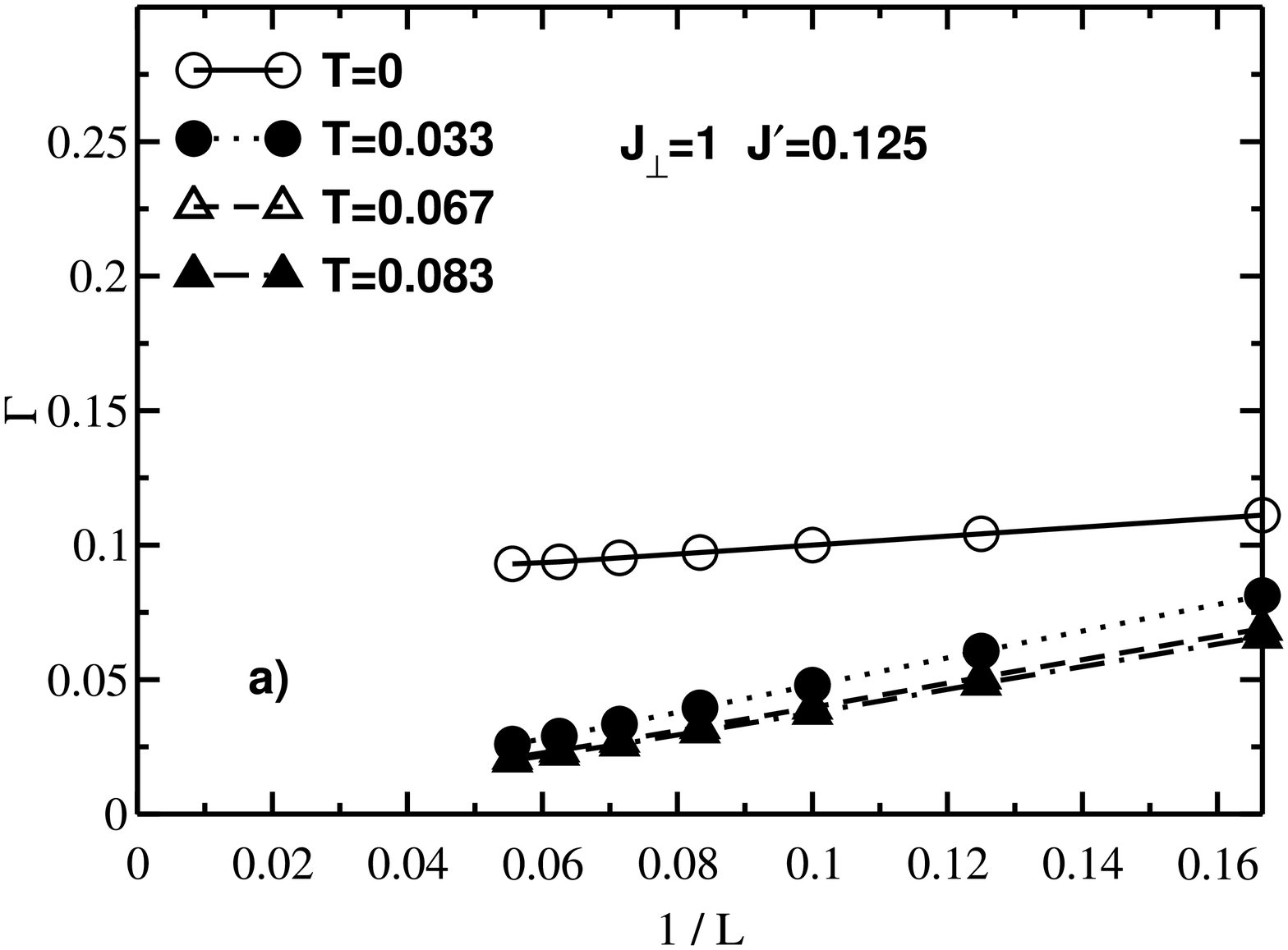}
\includegraphics[width=8cm]{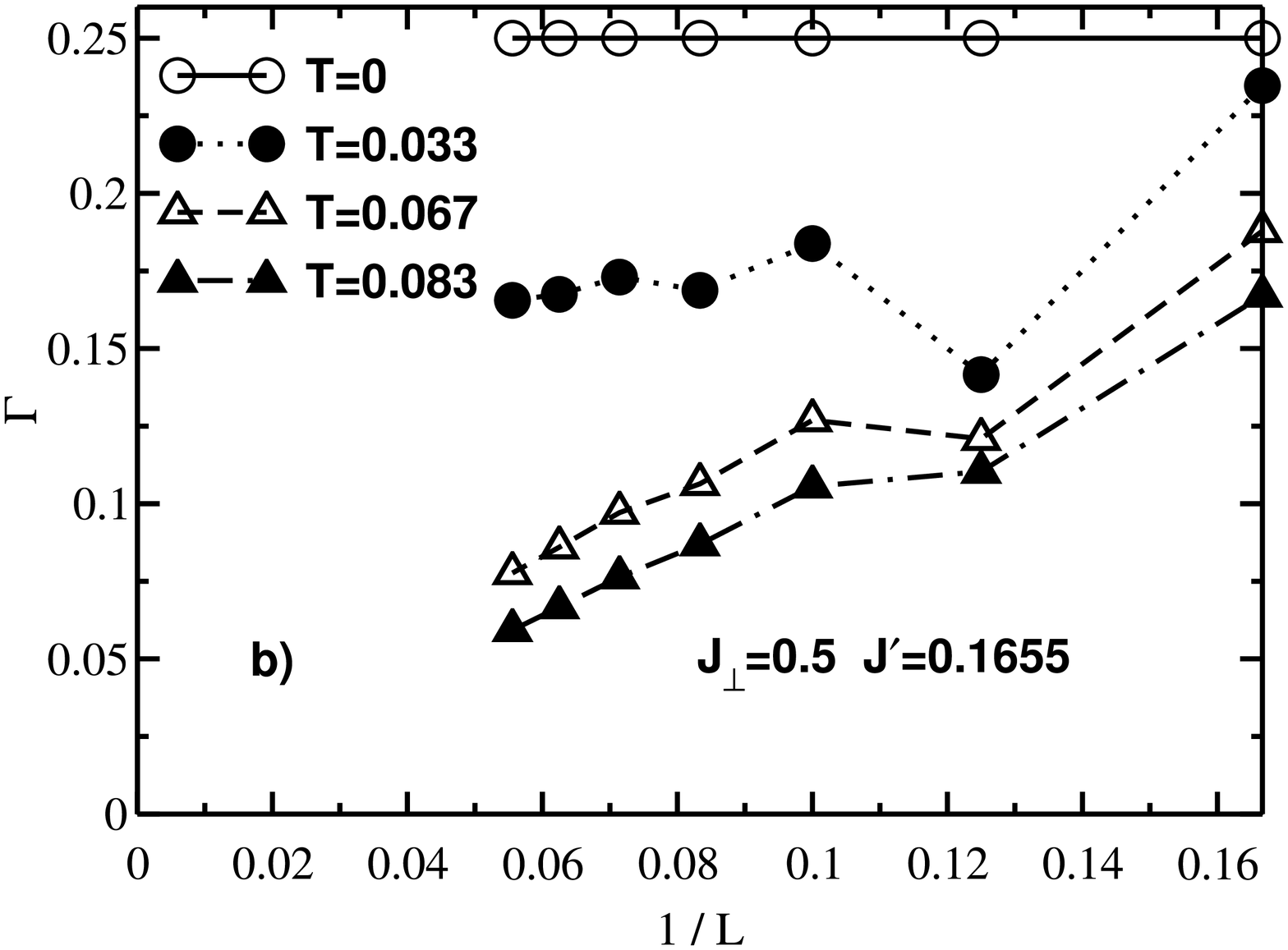}
}\vspace{0.9cm} \centerline{
\includegraphics[width=8cm]{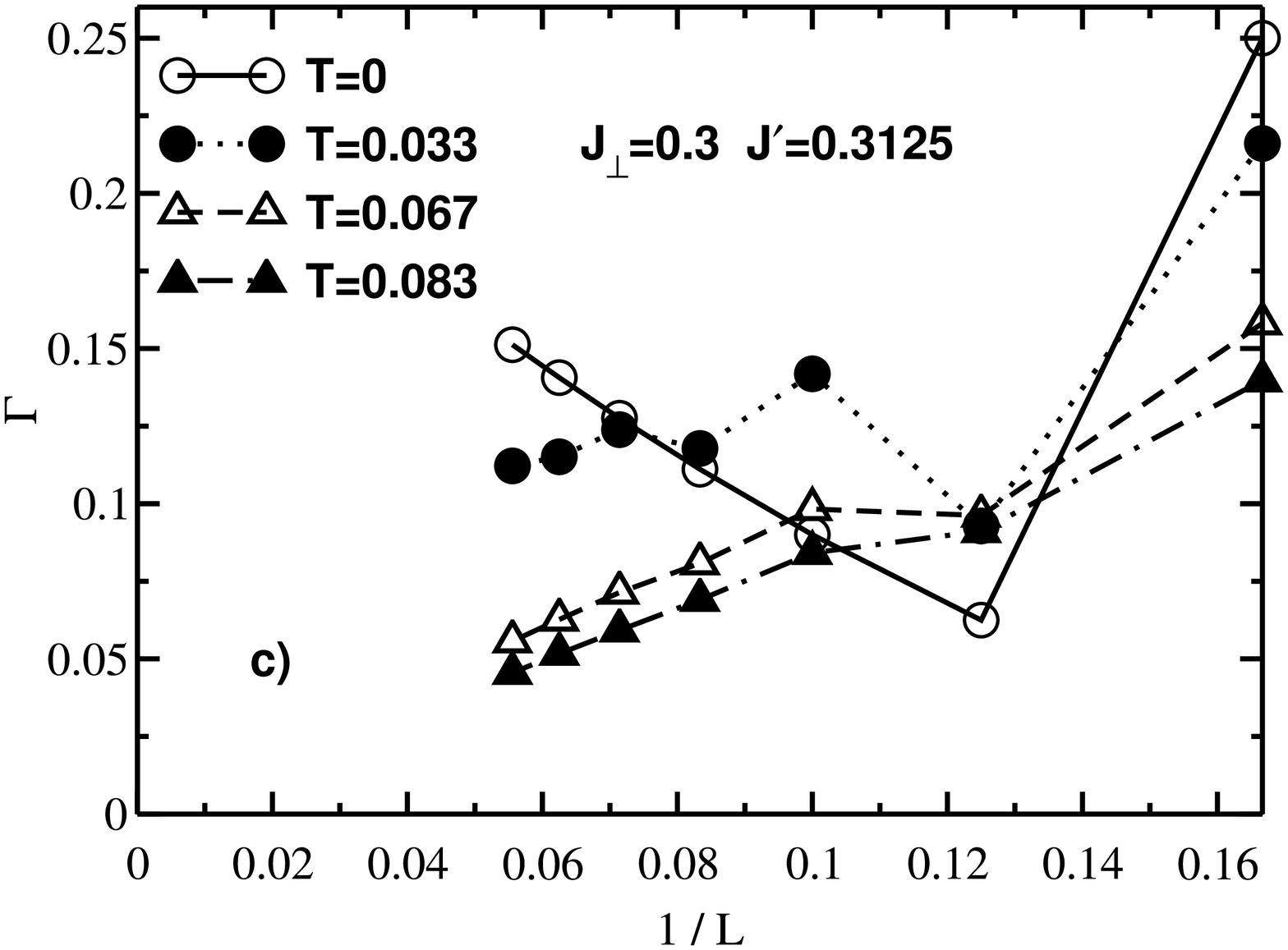}
\includegraphics[width=8cm]{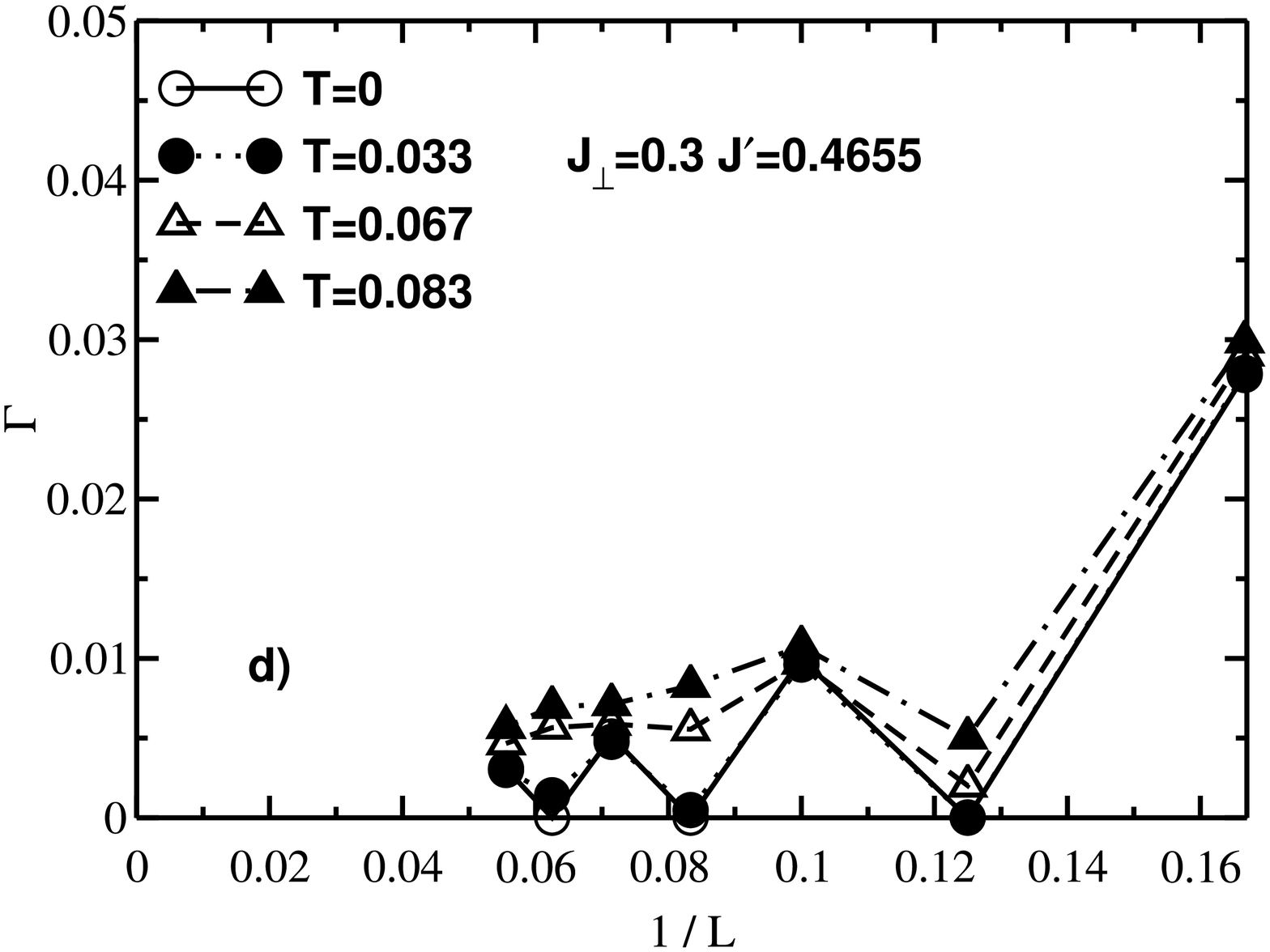}
} \caption{Finite-size scaling of the finite-temperature results for
different representative points on the phase diagram of
Fig.~\ref{fig1}. a) NE-I phase. b) NE-II phase. c), d) two sample
points from the \emph{weird} (W) phase, which at $T=0$ show quite
distinct finite-size scaling and at $T>0$ behave very similarly. All
energies are measured in units of $J_z$. All the data suggest that
the bulk system at $T>0$ is always ergodic.}\label{fig3}
\end{figure*}

\section{Finite-temperature results}\label{tn0}

At zero temperature, the degeneracy of the ground state is clearly
the source of non-ergodicity. At finite temperatures, all states
contribute to the dynamics of the system and we have to search for
some other source of non-ergodicity, if any. In fact, when the sum
in (\ref{Gammadef}) becomes an integral, the degree of degeneracy of
a state with energy $\varepsilon$ is given by the value of the
density of states at $\varepsilon$. Then, the result of the
integration is hard to guess. Once again, we have used finite-size
scaling in order to reveal the behavior in the bulk system. Ergodic
value of $\Gamma$ should now be calculated through the thermal
average:
\begin{equation}
   \Gamma^{erg} = \left (
   \frac{1}{Z} \sum_n e^{-\beta E_n}
   \langle
   n | S^z_i | n
   \rangle
   \right )^2.
\end{equation}
It follows that $\Gamma^{erg}$ is just zero for all Hamiltonian
couplings.

We have computed finite-size scaling at several finite temperatures
using ED. In Fig.~\ref{fig3}, four representative points from the
phase diagram in Fig.~\ref{fig1} are shown: the plots from panels a)
and b) belong to the former NE-I and NE-II phases, while those from
panels c) and d) belong to the W one. Finite-size scaling for the
ergodic phases (E-I and E-II) gives $\Gamma$ identically zero and is
not reported. Already at low temperatures ($T \sim 0.08 J_z$) and
small sizes $L \leq 18$, our calculations show a very clear
indication that, in the thermodynamic limit, $\Gamma$ is zero in the
whole phase diagram. In addition, we have found that $\Gamma$
approaches zero nearly proportional to $L^{-1}$ for finite systems.

These results can be analyzed from another point of view. As it is
stated by Suzuki \cite{Suzuki_71}, necessary and sufficient
condition for the ergodicity of the dynamics of an operator $A$
under the Hamiltonian $H$ is the orthogonality of $A$ to all the
integrals of motion of the system
\begin{equation} \label{ortho}
(A , H^i) = 0,
\end{equation}
where $\{H^i\}$ represents the complete set of integrals of motion
of the Hamiltonian $H$. $\{H^i\}$ form a linear space with metrics
given by the normalized density matrix
\begin{equation} \label{scal}
(H^i,H^j) = \frac1Z\sum_k e^{ -\beta E_k } H^i_{k,k} H^j_{k,k}
\equiv \langle H^i H^j \rangle,
\end{equation}
where $H^i$ and $H^j$ are two integrals of motion and the sum is
extended on all microscopic states of the system. Then, for a given
operator $A$, to be orthogonal to all integrals of motion of $H$
means either (i) to be off-diagonal with respect to $H$ or (ii) to
have a negligible number of diagonal matrix elements so that the
scalar products (\ref{ortho}) would go to zero in the thermodynamic
limit. It follows that the dynamics of an integral of motion is
surely non-ergodic, since it is surely not orthogonal to itself.

For finite systems, the response is practically always non-ergodic
as it is unlikely that (\ref{Gammadef}) is exactly zero. In fact,
for any fixed size, $\Gamma$ is non zero (see Fig.~\ref{fig3}).
Naively, one could expect that $S^z_\mathbf{i}$ has non-ergodic
dynamics also at finite temperatures since its average is
proportional to the one of $S^z_{tot}$, which is an integral of
motion. However, our numerical studies suggest that case (ii)
applies to this operator. It is worth noticing that this type of
behavior can be strongly dependent on the dimensionality of the
system. For instance, we can expect a non-ergodic dynamics also at
finite temperatures in higher dimensions, where a finite critical
temperature for the transition to the ferromagnetic phase exists.

As it results from our analysis, the question of ergodicity of a
spin system is closely connected to the presence of a finite
magnetic moment per site. It might seem, therefore, that since we
are dealing with a one-dimensional system, we could apply the
Mermin-Wagner theorem \cite{Mermin_66} and exclude the possibility
to have finite magnetization at $T \neq 0$. Actually, our model is
not isotropic so one of the conditions of Mermin-Wagner theorem is
not met. At any rate, even along the line $J_\perp = J_z$ the proof
of the above theorem involves the introduction of an infinitesimally
small magnetic field that would split the degeneracy and make
ergodic the system dynamics.

\section{conclusions}\label{conclusion}

Summarizing, we have studied the 1D anisotropic Heisenberg model
with next-nearest-neighbor exchange interaction and the ergodicity
of the operator $S^z_\mathbf{i}$ by means of the Lanczos and Exact
Diagonalization techniques. At zero temperature, we have constructed
the ergodicity phase diagram and found two distinct non-ergodic
regions in the thermodynamic limit. The borders of these regions and
the corresponding non-ergodic values of $\Gamma$ have been
determined by means of a finite-size scaling analysis of the Lanczos
data. The results of this analysis (i.e., critical couplings and
non-ergodic $\Gamma$ values, see (\ref{gcase})) can be used by
analytic methods that aim at computing physical properties and
response functions of this system. Our results for finite
temperatures suggest the ergodicity of the dynamics in the
thermodynamic limit in the whole range of couplings. It implies, on
the basis of Suzuki theorem \cite{Suzuki_71}, that the number of
diagonal matrix elements of $S^z_\mathbf{i}$ in the energy
representation is negligible in the thermodynamic limit. A
finite-size system, instead, remains non ergodic and the finite-$L$
corrections to the bulk value of $\Gamma$ are proportional to the
inverse volume of the system (i.e., to $L^{-1}$).

\acknowledgments One of the authors (E.P.) thanks Prof. A.A.
Nersesyan for the invaluable discussions on the issue.


\begin{thebibliography}{17}
\expandafter\ifx\csname
natexlab\endcsname\relax\def\natexlab#1{#1}\fi
\expandafter\ifx\csname bibnamefont\endcsname\relax
  \def\bibnamefont#1{#1}\fi
\expandafter\ifx\csname bibfnamefont\endcsname\relax
  \def\bibfnamefont#1{#1}\fi
\expandafter\ifx\csname citenamefont\endcsname\relax
  \def\citenamefont#1{#1}\fi
\expandafter\ifx\csname url\endcsname\relax
  \def\url#1{\texttt{#1}}\fi
\expandafter\ifx\csname urlprefix\endcsname\relax\def\urlprefix{URL
}\fi \providecommand{\bibinfo}[2]{#2}
\providecommand{\eprint}[2][]{\url{#2}}

\bibitem[{\citenamefont{Mancini and Avella}(2003)}]{Mancini_00}
\bibinfo{author}{\bibfnamefont{F.}~\bibnamefont{Mancini}} \bibnamefont{and}
  \bibinfo{author}{\bibfnamefont{A.}~\bibnamefont{Avella}},
  \bibinfo{journal}{Eur.~Phys.~J.~B} \textbf{\bibinfo{volume}{36}},
  \bibinfo{pages}{37} (\bibinfo{year}{2003}).

\bibitem[{\citenamefont{Mancini and Avella}(2004)}]{Mancini_04}
\bibinfo{author}{\bibfnamefont{F.}~\bibnamefont{Mancini}} \bibnamefont{and}
  \bibinfo{author}{\bibfnamefont{A.}~\bibnamefont{Avella}},
  \bibinfo{journal}{Adv.~Phys.} \textbf{\bibinfo{volume}{53}},
  \bibinfo{pages}{537} (\bibinfo{year}{2004}).

\bibitem[{\citenamefont{Suzuki}(1971)}]{Suzuki_71}
\bibinfo{author}{\bibfnamefont{M.}~\bibnamefont{Suzuki}},
  \bibinfo{journal}{Physica} \textbf{\bibinfo{volume}{51}},
  \bibinfo{pages}{277} (\bibinfo{year}{1971}).

\bibitem[{\citenamefont{Morita and Katsura}(1969)}]{Morita_69}
\bibinfo{author}{\bibfnamefont{T.}~\bibnamefont{Morita}} \bibnamefont{and}
  \bibinfo{author}{\bibfnamefont{S.}~\bibnamefont{Katsura}},
  \bibinfo{journal}{J.~Phys.~C.} \textbf{\bibinfo{volume}{2}},
  \bibinfo{pages}{1030} (\bibinfo{year}{1969}).

\bibitem[{\citenamefont{Mancini}(2005)}]{Mancini_05b}
\bibinfo{author}{\bibfnamefont{F.}~\bibnamefont{Mancini}},
  \bibinfo{journal}{Eur.~Phys.~J.~B} \textbf{\bibinfo{volume}{45}},
  \bibinfo{pages}{497} (\bibinfo{year}{2005}).

\bibitem[{\citenamefont{Bak et~al.}(2003)\citenamefont{Bak, Avella, and
  Mancini}}]{Bak_02a}
\bibinfo{author}{\bibfnamefont{M.}~\bibnamefont{Bak}},
  \bibinfo{author}{\bibfnamefont{A.}~\bibnamefont{Avella}}, \bibnamefont{and}
  \bibinfo{author}{\bibfnamefont{F.}~\bibnamefont{Mancini}},
  \bibinfo{journal}{Phys.~Stat.~Sol.~(b)} \textbf{\bibinfo{volume}{236}},
  \bibinfo{pages}{396} (\bibinfo{year}{2003}).

\bibitem[{\citenamefont{Tonegawa and Harada}(1987)}]{Tonegawa_87}
\bibinfo{author}{\bibfnamefont{T.}~\bibnamefont{Tonegawa}} \bibnamefont{and}
  \bibinfo{author}{\bibfnamefont{I.}~\bibnamefont{Harada}},
  \bibinfo{journal}{J.~Phys.~Soc.~Jpn.} \textbf{\bibinfo{volume}{56}},
  \bibinfo{pages}{2153} (\bibinfo{year}{1987}).

\bibitem[{\citenamefont{Harada et~al.}(1988)\citenamefont{Harada, Kimura, and
  Tonegawa}}]{Harada_88}
\bibinfo{author}{\bibfnamefont{I.}~\bibnamefont{Harada}},
  \bibinfo{author}{\bibfnamefont{T.}~\bibnamefont{Kimura}}, \bibnamefont{and}
  \bibinfo{author}{\bibfnamefont{T.}~\bibnamefont{Tonegawa}},
  \bibinfo{journal}{J.~Phys.~Soc.~Jpn.} \textbf{\bibinfo{volume}{57}},
  \bibinfo{pages}{2779} (\bibinfo{year}{1988}).

\bibitem[{\citenamefont{Nomura and Okamoto}(1993)}]{Nomura_93}
\bibinfo{author}{\bibfnamefont{K.}~\bibnamefont{Nomura}} \bibnamefont{and}
  \bibinfo{author}{\bibfnamefont{K.}~\bibnamefont{Okamoto}},
  \bibinfo{journal}{J.~Phys.~Soc.~Jpn.} \textbf{\bibinfo{volume}{62}},
  \bibinfo{pages}{1123} (\bibinfo{year}{1993}).

\bibitem[{\citenamefont{Nomura and Okamoto}(1994)}]{Nomura_94}
\bibinfo{author}{\bibfnamefont{K.}~\bibnamefont{Nomura}} \bibnamefont{and}
  \bibinfo{author}{\bibfnamefont{K.}~\bibnamefont{Okamoto}},
  \bibinfo{journal}{J. Phys. A} \textbf{\bibinfo{volume}{27}},
  \bibinfo{pages}{5773} (\bibinfo{year}{1994}).

\bibitem[{\citenamefont{Hirata and Nomura}(2000)}]{Hirata_00}
\bibinfo{author}{\bibfnamefont{S.}~\bibnamefont{Hirata}} \bibnamefont{and}
  \bibinfo{author}{\bibfnamefont{K.}~\bibnamefont{Nomura}},
  \bibinfo{journal}{Phys.~Rev.~B} \textbf{\bibinfo{volume}{61}},
  \bibinfo{pages}{9453} (\bibinfo{year}{2000}).

\bibitem[{\citenamefont{Haldane}(1982)}]{Haldane_82}
\bibinfo{author}{\bibfnamefont{F.~D.~M.} \bibnamefont{Haldane}},
  \bibinfo{journal}{Phys.~Rev.~B} \textbf{\bibinfo{volume}{25}},
  \bibinfo{pages}{R4925} (\bibinfo{year}{1982}).

\bibitem[{\citenamefont{Shastry and Sutherland}(1981)}]{Shastry_81}
\bibinfo{author}{\bibfnamefont{B.~S.} \bibnamefont{Shastry}} \bibnamefont{and}
  \bibinfo{author}{\bibfnamefont{B.}~\bibnamefont{Sutherland}},
  \bibinfo{journal}{Phys.~Rev.~Lett.} \textbf{\bibinfo{volume}{47}},
  \bibinfo{pages}{964} (\bibinfo{year}{1981}).

\bibitem[{\citenamefont{Majumdar}(1970)}]{Majumdar_70}
\bibinfo{author}{\bibfnamefont{C.~K.} \bibnamefont{Majumdar}},
  \bibinfo{journal}{J.~Phys.~C.} \textbf{\bibinfo{volume}{3}},
  \bibinfo{pages}{911} (\bibinfo{year}{1970}).

\bibitem[{\citenamefont{Frahm}(1992)}]{Frahm_92}
\bibinfo{author}{\bibfnamefont{H.}~\bibnamefont{Frahm}}, \bibinfo{journal}{J.
  Phys. A} \textbf{\bibinfo{volume}{25}}, \bibinfo{pages}{1417}
  (\bibinfo{year}{1992}).

\bibitem[{\citenamefont{Cyr et~al.}(1996)}]{Cyr_96}
\bibinfo{author}{\bibfnamefont{S.~L.} \bibnamefont{Cyr}} \bibnamefont{et~al.},
  \bibinfo{journal}{J.~Phys.:~Condens.~Matter} \textbf{\bibinfo{volume}{8}},
  \bibinfo{pages}{4781} (\bibinfo{year}{1996}).

\bibitem[{\citenamefont{Mermin and Wagner}(1966)}]{Mermin_66}
\bibinfo{author}{\bibfnamefont{N.~D.} \bibnamefont{Mermin}} \bibnamefont{and}
  \bibinfo{author}{\bibfnamefont{H.}~\bibnamefont{Wagner}},
  \bibinfo{journal}{Phys.~Rev.~Lett.} \textbf{\bibinfo{volume}{17}},
  \bibinfo{pages}{1133} (\bibinfo{year}{1966}).

\end{thebibliography}

\end{document}